\documentclass[pra, aps, twocolumn]{revtex4-1}
\usepackage{graphicx,graphics,psfrag,amsmath,calc,mathtools}
\usepackage{epsfig, bm}
\usepackage{color,comment}
\usepackage{float}

\def\ba{\begin{eqnarray}}
\def\ea{\end{eqnarray}}

\usepackage{tikz,fp}
\usepackage{tikz-cd}
\usetikzlibrary{arrows}
\usetikzlibrary{intersections}
\usetikzlibrary{shapes.geometric}
\usetikzlibrary{decorations.pathmorphing, patterns,shapes,fixedpointarithmetic}
\usetikzlibrary{decorations.markings}

\pgfdeclarepatternformonly{south west lines}{\pgfqpoint{-0pt}{-0pt}}{\pgfqpoint{3pt}{3pt}}{\pgfqpoint{3pt}{3pt}}{
	\pgfsetlinewidth{0.4pt}
	\pgfpathmoveto{\pgfqpoint{0pt}{0pt}}
	\pgfpathlineto{\pgfqpoint{3pt}{3pt}}
	\pgfpathmoveto{\pgfqpoint{2.8pt}{-.2pt}}
	\pgfpathlineto{\pgfqpoint{3.2pt}{.2pt}}
	\pgfpathmoveto{\pgfqpoint{-.2pt}{2.8pt}}
	\pgfpathlineto{\pgfqpoint{.2pt}{3.2pt}}
	\pgfusepath{stroke}}

\tikzset{
	mid arrow/.style={postaction={decorate,decoration={
				markings,
				mark=at position .575 with {\arrow{stealth}}
	}}},
	near arrow/.style={postaction={decorate,decoration={
				markings,
				mark=at position .275 with {\arrow{stealth}}
	}}},
	far arrow/.style={postaction={decorate,decoration={
				markings,
				mark=at position .800 with {\arrow{stealth}}
	}}},
	snake arrow/.style={fixed point arithmetic, decorate, decoration={snake,amplitude=2pt, segment length=11pt},postaction={decoration={markings,mark=at position 0.625 with {\arrow{stealth}}},decorate}},
}

\usepackage{color}

\usetikzlibrary {decorations.fractals}
\usetikzlibrary { decorations.pathmorphing, decorations.pathreplacing, decorations.shapes, }

\begin{document}
\date{\today}

\title{A Hydrodynamical Description of Bose Liquid with Fractional Exclusion Statistics}

\author{Zhaoyu Fei$^{1}$}
\author{Yu Chen$^1$}   \thanks{ychen@gscaep.ac.cn}
\affiliation{$^1$Graduate School of China Academy of Engineering Physics, Beijing 100193, China}
\date{\today}

\begin{abstract}
Hydrodynamical systems are usually taken as chaotic systems with fast relaxations. It is counter intuitive for ``ideal" gas to have a hydrodynamical description.
We find that a hydrodynamical model of one-dimensional $|\Phi|^6$ theory shares the same ground state density profile, density-wave excitation, as well as the similar dynamical and statistical properties
with the Calogero-Sutherland model in thermodynamic limit when their interaction strengths matches each other.  The interaction strength $g_0$ in $|\Phi|^6$ theory is then the index of fractional statistics.
Although the model is interacting in Bose liquid sense, but it shows integrability with periodical coherent evolution. We also discussed the fractional statistics emerges from the $|\Phi|^6$ theory. 
\end{abstract}

\maketitle

%
\section{Introduction}

In conventional wisdom of hydrodynamics, the existence of the hydrodynamical equations are based on the rapid relaxation of the system~\cite{Landau}.
Usually this assumption is well satisfied in chaotic systems such like the interacting Bose gases.
However, there are some interesting exceptions, such as the SGZ solutions which shows integrable motions in hydrodynamics~\cite{shi2021}, where it shows hydrodynamics equations can also emerge for ``ideal" systems. These integrability are shown in the examples of two-dimensional superfluids~\cite{paris,qi20}. On the other hand, we also notice that the ideal anyons satisfying fractional exclusion statistics which can be described by interacting bosons or fermions ~\cite{Haldane,wu1994}.
Naturally, one question is raised on whether the dynamics of ideal anyons can be treated  by specific integrable hydrodynamical theory. In this article, we are going to explore this problem.

The Calogero-Sutherland (CS) model~\cite{csm1969,sutherland246} describes trapped particles with $1/r^2$ long-range two-body interactions,
\ba
\hat{H}_{\rm CS}=\sum_{j=1}^{N}\left(\frac{-1}{2}\partial_{x_j}^2+\frac{1}{2}\omega_{\rm cs}^2 x_j^2\right)+\sum_{i<j}\frac{g_{\rm cs}(g_{\rm cs}-1)}{|x_i-x_j|^2},
\ea
where $\partial_{x_j}=\partial/\partial x_j$, $N$ is the total particle number, $\omega_{\rm cs}$ the trap frequency and $g_{\rm cs}$ a dimensionless interaction strength. The particle mass $m$ and $\hbar$ are taken as 1. This model is exactly solvable and one can prove that the excitations of CS model obey fractional exclusion statistics~\cite{murthy1994}. Such statistics was first introduced by Haldane, and Yongshi Wu~\cite{Haldane, wu1994}. Here, we stress that the parameter $g_{\rm cs}$ is the index labelling the fractional statistics. When $g_{\rm cs}=0$, the system is ideal Bosons. And when $g_{\rm cs}=1$, the system is  impenetrable Bosons, which can be
mapped to spinless ideal Fermions in terms of the theorems in~\cite{girardeau1960}.

We find the one-dimensional $|\Phi|^6$ Bose liquid theory shares the same ground state density profile, density-wave excitation   as well as the similar dynamical and statistical properties as them of the CS model.   The trapped $|\Phi|^6$ theory can be given as
\ba
\label{energy}
H=\frac{1}{2}\int_{-\infty}^{+\infty}\!\!\!\!\!\!\!\mathrm dx\left[\left|\frac{\partial\Phi}{\partial x}\right|^2\!\!\!+\omega_0^2x^2 |\Phi|^2+\frac{g_0^2\pi^2}{3}|\Phi|^6\right]\!\!,
\ea
where $\Phi(x)$ is a one-dimensional complex field. $\omega_0$ is the trap frequency and $g_0$ is the three-body contact interaction strength. Remarkably, we find $g_0$ and $\omega_0$ matches $g_{\rm cs}$ and $\omega_{\rm cs}$ exactly. More precisely, it means that both the ground state density profile of the CS model and it of the $|\Phi|^6$ model share the same Wigner semicircle distribution in the thermodynamic limit~\cite{wigner1955}. Here, the thermodynamic limit means that $N\to\infty, \omega\to 0$ while keep $N\omega$ as a constant~\cite{dalfovo1999}. The excitation frequencies are equal spacing. The period of the density wave excitation is interaction strength independent and trap frequency dependent only.
The direct consequence of this property is the coherence in time evolution. We also find that the $|\Phi|^6$ theory satisfies the fractional exclusion statistics whose statistical parameter is $g_0$ corresponding to $g_{\rm cs}$ in the CS model. It is worth mentioning that the connection between the two models has been implicitly revealed by using the collective-field method~\cite{Andric1988, Sen1997}.

In the following, we will present the correspondence between the CS model and trapped $|\Phi|^6$ theory from the dynamical symmetry, ground state profile, excitation spectra, dynamical behavior, and statistical properties respectively. These correspondence are verified analytically in the thermodynamic limit and numerically checked for a finite-$N$ case.

\section{The Correspondence Between $|\Phi|^6$ theory and the  Calogero-Sutherland model}

\subsection{$\mathrm{SU}(1, 1)$ dynamical symmetry}

Previously, it is discovered that for hamiltonian of the form as $\hat{H}=-\frac{1}{2}\sum_{i} \partial_i^2 +V(x_1,\cdots, x_N)+  \frac{1}{2}\sum_i \omega_0^2 x_i^2 $, the system has $SU(1,1)$ dynamical symmetry as long as $V(x_1,\cdots, x_N)$ satisfying $V(\lambda x_1,\cdots,\lambda x_N)=\lambda^{-2}V(x_1,\cdots, x_N)$. A simple review of this symmetry can be checked by the commutative relation among $\hat{H}_0=-\frac{1}{2}\sum_{i} \partial_i^2  +V(x_1,\cdots, x_N)$, $\hat{I}=\frac{1}{2}\sum_{i}\omega^2_0 x_i^2$, and $\hat{Q}=-\frac{i}{2}\sum_{i}(\partial_i x_i+x_i \partial_i)$, i. e.,
\ba
[\hat{Q},\hat{H}_0]=2i\hat{H}_0,\hspace{4ex}[\hat{Q},\hat{I}]=-2i \hat{I},\nonumber
\ea
\ba
[\hat{I},\hat{H}_0]=i\omega_0^2 \hat{Q}.\nonumber
\ea
By introducing $L_1=(\hat{H}_0-\hat{I})/2\omega_0$, $L_{3}=\hat{H}/2\omega_0$, and $L_2=\hat{Q}/2$, we have
\ba
[L_1,L_2]=-iL_3,\hspace{3ex}[L_2,L_3]=iL_1,\hspace{3ex}[L_3,L_1]=iL_2.\nonumber
\ea
Clearly, these three operators form  $\mathfrak{su}(1,1)$ algebra. 
We can check that both of the CS model and the trapped $|\Phi|^6$ theory have this scale property. For the $|\Phi|^6$ theory, its microscopic first quantization interaction term is a three-body contact interaction
\ba
V(x_1,\cdots, x_N)=g_0^2\pi^2\sum_{i<j<k}\delta(x_i-x_j)\delta(x_j-x_k).
\ea
The dynamical symmetry is easily checked by noticing $\delta(\lambda x)=\delta(x)/|\lambda|$, which is also inherited by the trapped $|\Phi|^6$ theory.
\begin{figure*}[t]
\centering
\includegraphics[width=15cm]{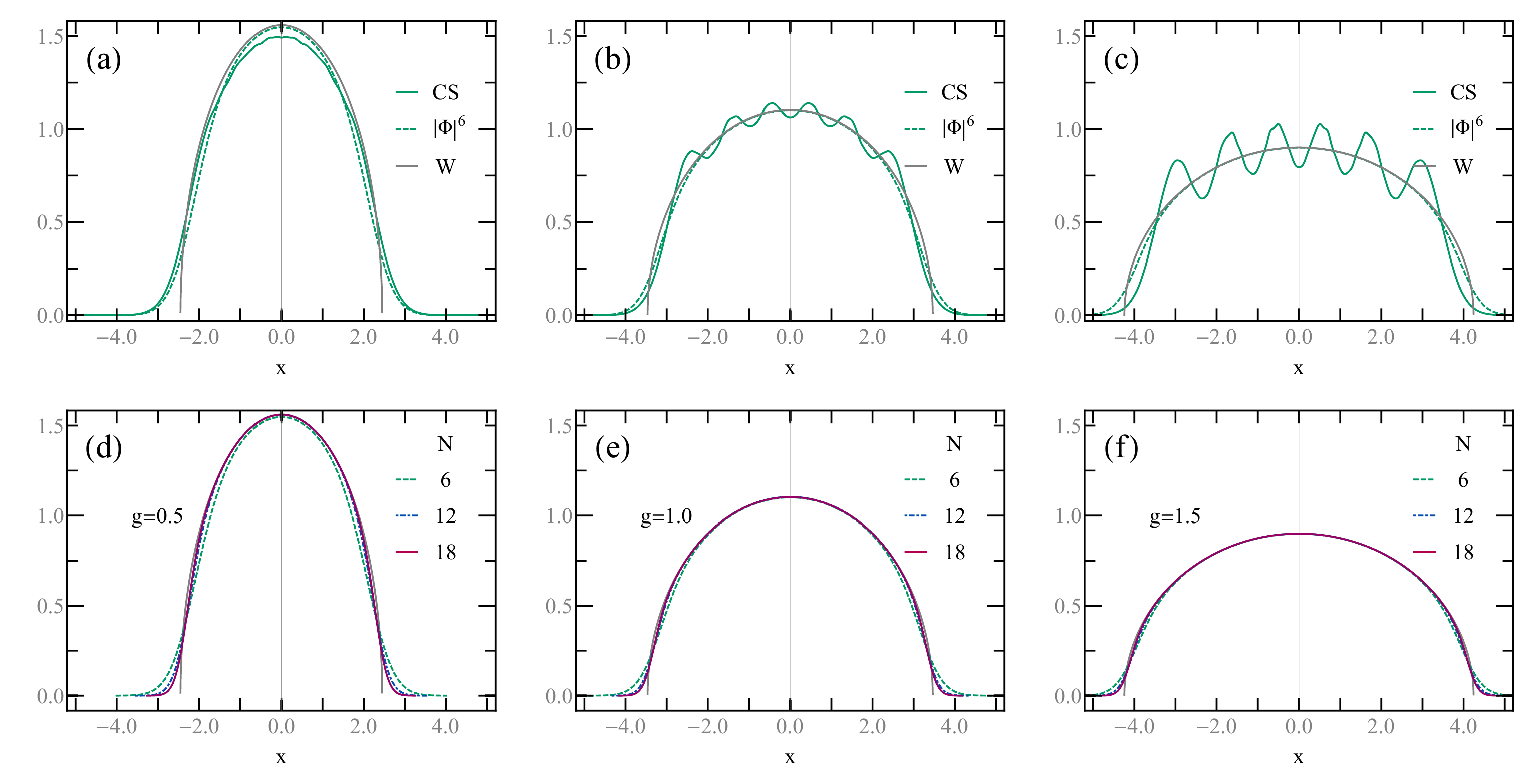}
\caption{In (a), (b) and (c) we show the ground state density profile of CS model (solid line, short for CS), $|\Phi|^6$ theory (dashed line, short for $|\Phi|^6$) and Wigner semicricle distribution (Gray line, short for W) for $N=6$. $g=g_{\rm cs}=g_0=0.5$, $g=g_{\rm cs}=g_0=1.0$ and $g=g_{\rm cs}=g_0=1.5$ in (a), (b) and (c) respectively. In (d), (e) and (f), we show how the ground state density profile of the $|\Phi|^6$ theory approaches Wigner semicircle distribution in the thermodynamic limit. The interaction strength is taken as $0.5$ in (d), $1.0$ in (e) and $1.5$ in (f). $N=6, 12, 18$ respectively. For comparison, the ground state density profile for $N=12$ and $N=18$ are rescaled such that their radius $R=\sqrt{2g_{0}N/\omega_0}$ and the area under the curve are the same as them in the $N=6$ case. }
\label{FigGS}
\end{figure*}

The direct result for this dynamical symmetry is a $2\omega_0$ frequency breathing mode in the trap~\cite{pitaev1997}. We also stress that the symmetry property is correct for any particle number $N$. But the symmetry alone can not ensure the correspondence between the two systems. Actually we can see the deviation between these two theories at large $g_0=g_{\rm cs}$ with a finite $N$, where the long-range interaction in the CS model induce a Wigner crystal~\cite{wigner} while this density order is clearly absent in the $|\Phi|^6$ theory.

\subsection{  ground state density profile}

In this section, we first verify that the ground state density profile in the CS model and trapped $|\Phi|^6$ theory are identical in the thermodynamic limit. Then, for a finite $N$, we present the numerical results of the ground state density $\rho^{\rm CS}_{\rm g}(x)=|\Psi_{\rm g}(x)|^2$, $\rho_{\rm g}(x) $ for the CS model and the $|\Phi|^6$ theory respectively.

In the $|\Phi|^6$ theory, the ground state density profile can be obtained by an equivalent hydrodynamical description, where the density and velocity field obey the following equations,
\ba
\frac{\partial}{\partial t}\rho+\frac{\partial}{\partial x}(\rho v)=0\label{Hydro01},
\ea
\ba
\frac{\partial v}{\partial t}\!+\!\frac{\partial}{\partial x}\left(\frac{1}{2}v^2\!+\!\frac{1}{2}\omega_0^2 x^2\!+\!\frac{g_0^2\pi^2\rho^2}{2}\!+\!P_Q\!\right)\!=0\label{Hydro02}.
\ea
Here, $\rho=\rho(x,t)\equiv|\Phi(x,t)|^2$ is the density field, and $v=v(x,t)\equiv\left(\Phi^*\partial_x\Phi-\partial_x\Phi^*\Phi\right)/2i\rho(x,t)$ is the velocity field. $P_Q=- \sqrt{\rho(x,t)}\partial^2_x\sqrt{\rho(x,t)}/2$ is the so-called quantum pressure. When the trap varies slowly on the scale of the inter-particle spacing, the density profile $\rho$ becomes smooth and the quantum pressure $P_Q$ can be neglected, which is called the Thomas-Fermi (TF) approximation. Because in ground state, the velocity field is zero, $\partial_t\rho=0$, then it requires
\ba
\frac{\partial}{\partial x}\left( \omega_0^2 x^2+ g_0^2\pi^2\rho^2 \right)=0.
\ea
Therefore, the ground state density profile is
\ba
\rho_{\rm g}(x)= \frac{\omega_0}{\pi g_0}\sqrt{\frac{2g_0 N}{\omega_0}-x^2 }\theta\left(\sqrt{\frac{2g_0N}{\omega_0}}-|x|\right),\label{Phi6Ground}
\ea
where $N=\int {\mathrm dx}\rho(x)$ is the total particle number, $\theta(x)$ denotes the Heaviside step function.
When $g_0=1$, such an equation is the ground state profile of spinless ideal fermions in a harmonic trap, which has been studied in Ref.~\cite{eugene2000}
Although this result is obtained under TF approximation, we  see the semicircle law is quickly reached for finite $N$ with our numerical simulations.

Next, we solve the ground state density profile for the CS model. The expression of the ground state of the CS model is given as
\ba
\Psi_{\rm g} (\{x_j\})=\sqrt{C}\prod_{1\leq i<j\leq N}|x_i-x_j|^{g_{\rm cs}} e^{-\frac{\omega_{\rm cs}}{2}\sum_jx_j^2},
\ea
with the ground state energy $E_g=\omega_{\rm cs}[N/2+g_{\rm cs}N(N-1)/2]$,
where 
$x_j$ are the positions of particles, and
$C$ is the normalization factor, whose inverse is
\ba
C^{-1}=\left(\frac{\omega_{\rm cs}}{2}\right)^{- E_g/\omega_{\rm cs} }\left(\frac{\pi}{2}\right)^{N/2}\prod_{j=1}^{N}\frac{\Gamma(1+j g_{\rm cs})}{\Gamma(1+g_{\rm cs})}.
\ea
The density profile is obtained by $\rho^{\rm CS}_{\rm g}(x)=\int \prod_{j=2}^N \mathrm dx_j |\Psi_g (x, x_2,\cdots, x_N)|^2$. As the wave function itself is just the joint probability of the eigenvalues for random Gaussian matrices ensemble, the density function can be viewed as the charge distribution of log-potential Coulomb gas~\cite{forrester2010}. In the thermodynamic limit, the solution of the density profile is obtained
\ba
\rho^{\rm CS}_{\rm g}(x)= \frac{\omega_{\rm cs}}{\pi g_{\rm cs}}\sqrt{\frac{2g_{\rm cs}N}{\omega_{\rm cs}}-x^2}\theta\left(\sqrt{\frac{2g_{\rm cs}N}{\omega_{\rm cs}}}-|x|\right).\label{CSGround}
\ea
By comparing Eq.~(\ref{Phi6Ground}) and Eq.~(\ref{CSGround}), we find that these two density profiles are exactly the same when we match the frequencies and the interaction strength.
It is worth mentioning that $\rho_{\rm g}(x)$  also obtained from the zero-temperature distribution of the occupation number $n(\epsilon)$  for   anyons satisfying fractional exclusion statistics~\cite{murthy1994}:
\begin{gather}
 \begin{split}
n(\epsilon)=g_0^{-1}\theta(g_0N\omega_0-\epsilon)
 \end{split}
\end{gather}
with a semiclassical single-particle energy  $\epsilon=p^2/2+\omega_0^2x^2/2$. Then, $\rho_{\rm g}(x)$ is given by $ \rho_{\rm g}(x)=(2\pi)^{-1}\int\mathrm dp n(\epsilon)$.

In the following, we are going to show the numerical results for ground state density profile for the CS model and for the $|\Phi|^6$ theory. Here and in the following figures, we set $\omega_{\rm cs}=\omega_0=1$. For the CS model, we use the Monte-Carlo integration. Here we keep $x_1=x$, and we reparametrize $x_{i=2}$ to $x_{i=N}$ on a $(N-1)$-sphere $\sum_{i=2}^Nx_i^2=r^2$. The measure can be then written as $\prod_{j=2}^{N}\mathrm dx_i=r^{N-2}\mathrm  dr \mathrm d\Omega_{N-1}$ with $\Omega_{N-1}$ denoting the solid angle of the $(N-1)$-sphere. The ground state density profile is then expressed as
\ba
\rho_{\rm g}^{\rm CS}(x)&=&\int r^{N-2+g_{\rm cs}(N-1)(N-2)}e^{-\omega_{\rm cs} r^2}\mathrm dr\int \mathrm d\Omega_{N-1}\\
&&\prod_{j=2}^{N}|x\!-\!x_j(r,\Theta)|^{2g_{\rm cs}}\!\!\!\!\!\!\!\prod_{2\leq i<j\leq N}\!\!\!\!\!\!\!\left|\frac{x_i(r,\Theta)\!-\!x_j(r,\Theta)}{r}\right|^{2g_{\rm cs}}.\nonumber
\ea
Here $\Theta$ represent all the angle parameters of the $(N-1)$-sphere. 
From the expression,  the integrant is semi-positive definite, thus the multi-dimensional integral is very suitable for Monte-Carlo integration. Meanwhile, because of the measure and the Gaussian distribution in wave-function, the major contribution of above integration comes from very narrow region of $r=0$. The concentration of measure in  angular part due to L\'{e}vy's Lemma~\cite{Ledoux} can further reduce the complexity in integration. With all these helps, the ground state density profile of CS model for $N=3$, $N=6$ can be obtained.

On the other hand, we carry out a numerical calculation of $|\Phi|^6$ theory as a non-linear Schr\"{o}dinger equation. A cutoff in real space is taken as $X_{\rm max}$, and a boundary condition is taken as $\Phi(x=X_{\rm max})=\Phi(x=-X_{\rm max})$. To solve the ground state of the $|\Phi|^6$ theory, we turn on an evolution in imaginary time, where the ground state is the fixed point of this flow. 

The numerical results are summarized in Fig.~\ref{FigGS}. In Fig.~\ref{FigGS}(a), (b) and (c), we have shown the ground state density profile of CS model, the $|\Phi|^6$ theory and the semicircle law   with different interaction strength $g_{\rm cs}=g_0=0.5$, $1$ and $1.5$. For $g_{\rm cs}>1$, the density wave order becomes prominent. This is the tendency for Wigner crystal because of the long-range nature of $r^{-2}$ interaction. For $g_{\rm cs}=g_0<1$, the ground state density profile of CS model and the $|\Phi|^6$ model is very close to each other even for $N=6$. At the same time,  in Fig.~\ref{FigGS}(d), (e) and (f), the ground state density profile of the $|\Phi|^6$ theory  approaches the semicircle distribution as $N$ increases. Actually, the semicircle law is already a very good approximation for $N=6$.

\subsection{ The Correspondence in density-wave excitations and the coherent evolution}

In the following, we are going to connect the density-wave excitation of the $|\Phi|^6$ theory and the CS model in the thermodynamic limit. By comparison, we  find that the density-wave excitations of the two models share the same expression. 
Then, we  check the finite-$N$ effect  by full numerical calculations of the time evolution of the excitations. We analytically prove the correspondence between the density-wave excitations for any excited modes where $g_{\rm cs}=g_0=1$, or for the second excited mode where  $g_{\rm cs}=g_0$ is arbitrary.
A guess for the correspondence between the density-wave excitations of the two models for any excited modes and for arbitrary $g_{\rm cs}=g_0$  is made and is checked with finite $N$. From the numerical results we obtained, the guess seems to be hold in the thermodynamic limit.
\begin{figure}[t]
\centering
\includegraphics[width=8.5cm]{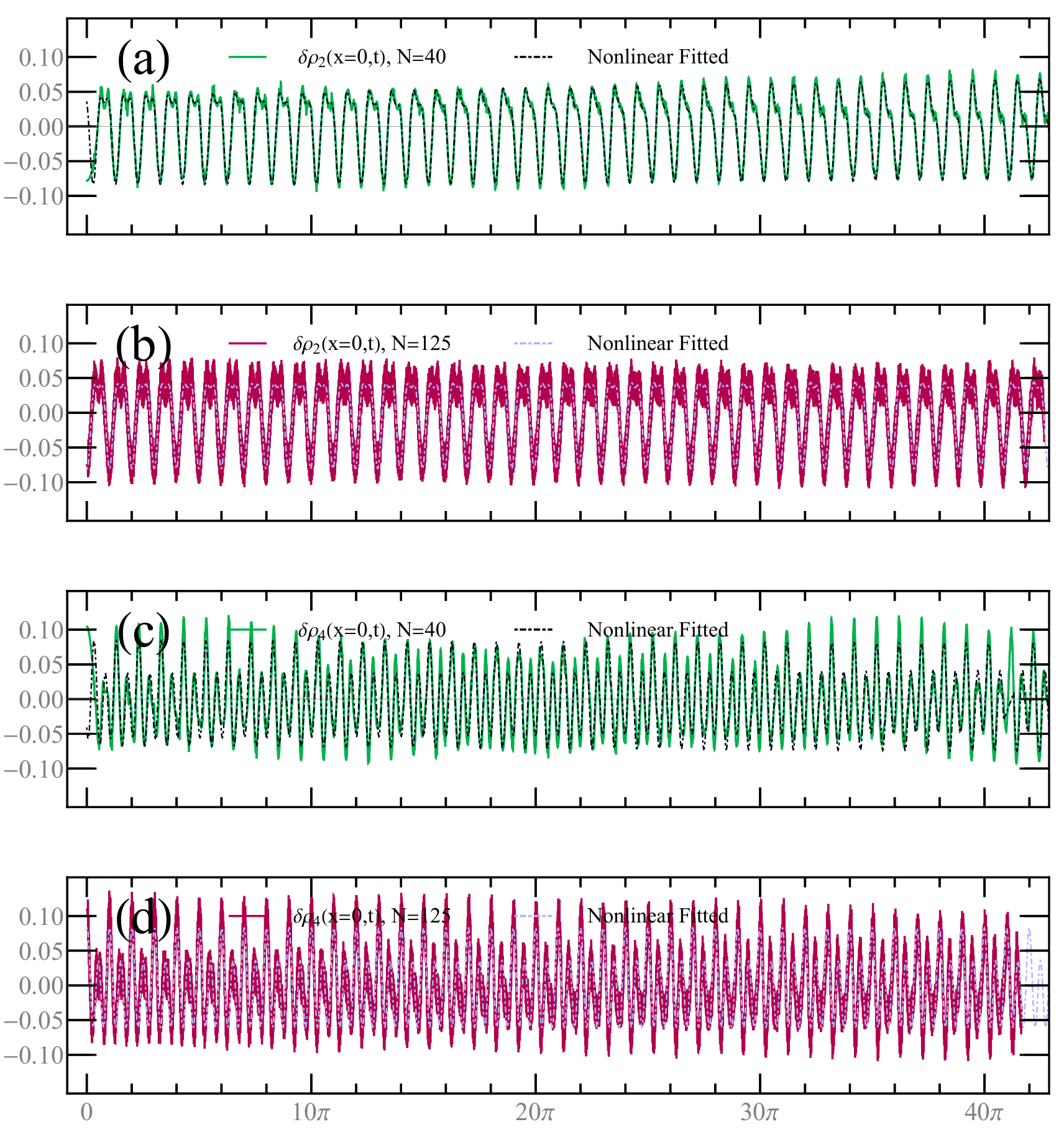}
\caption{In (a) and (c), we show the time evolution of $\delta\rho_n(x=0, t)$ for $n=2$ and $n=4$. The particle number $N=40$. In (b) and (d), we show the time evolution of $\delta\rho_n(x=0, t)$ for $n=2$ and $n=4$. The particle number $N=125$. All interaction strength are fixed at $g_0=\sqrt{2}$. The dotdashed lines are nonlinear fittings of the curves. The ansatz is taken as $0.1\sum_{n=2,4} a_n\cos(n\bar{\omega}_n+\varphi_n)$, $a_2$, $a_4$, $\bar{\omega}_2$, $\bar{\omega}_4$, $\varphi_2$ and $\varphi_4$ are fitting parameters. The fitting parameters are (a) $a_2=0.587$, $a_4=0.245$, $\bar{\omega}_2=1.006$, $\bar{\omega}_4=1.003$, $\varphi_2=1.35$, $\varphi_4=-0.465$; (b) $a_2=-0.584$, $a_4=-0.221$, $\bar{\omega}_2=1.00439$, $\bar{\omega}_4=1.00415$, $\varphi_2=-0.099$, $\varphi_4=-0.202$; (c)  $a_2=-0.584$, $a_4=-0.2208$, $\bar{\omega}_2=1.00439$, $\bar{\omega}_4=1.00415$; (d) $a_2=0.2358$, $a_4=0.5907$, $\bar{\omega}_2=1.0007$, $\bar{\omega}_4=1.0006$. }
\label{Exs}
\end{figure}

First of all, let us focus on the $|\Phi|^6$ theory. Here our starting point is the hydrodynamical equation Eq.~(\ref{Hydro01}) and Eq.~(\ref{Hydro02}).  Following the procedure in Refs.~\cite{string1996,menotti2002}, we linearize the density $\rho=\rho_0+\delta\rho$ (correspondingly $v=\delta v$) with a perturbation $\delta\rho$ obeying the linear expansion of Eqs.~(\ref{Hydro01}), (\ref{Hydro02})
\ba
\frac{\partial^2 \delta\rho}{\partial t^2}-\omega_0^2\frac{\partial}{\partial x}\left[\sqrt{R^2-x^2}\frac{\partial}{\partial x}(\sqrt{R^2-x^2}\delta\rho)\right]=0.
\ea
where $R=\sqrt{2g_0 N/\omega_0}$ is the radius of the semicircle distribution in the thermodynamic limit. 
The boundary conditions come from the absence of the particle flow at $x=\pm R$~\cite{imam2006}
\ba
\rho_0\delta v|_{x=\pm R}=0.
\ea
Then, by performing the transformation $\vartheta=\arccos(x/R)$ ($0\leq \vartheta\leq \pi$), $\delta\tilde{\rho}=\sqrt{R^2-x^2}\delta\rho,\delta \tilde{v}=\rho_0\delta v$, we obtain a standard wave function
\ba
\label{e16}
\frac{\partial^2 \delta\tilde{\rho}}{\partial t^2}-\omega^2\frac{\partial^2\delta\tilde{\rho}}{\partial \vartheta^2}=0,
\ea
with the Neumann boundary condition
\ba
\left.\frac{\partial \delta\tilde{\rho}}{\partial \vartheta}\right|_{\vartheta=0,\pi}=0.
\ea
Thus, the solution of $\delta\tilde{\rho}$ and $\delta \tilde{v}$  consists of Fourier series
\ba
\label{e18}
\delta\tilde{\rho}_n(\vartheta,t)&=&A_0\!+\!\!\sum_{n=1}^{\infty}\!A_n\cos(\omega_n t+\varphi_n)\cos(n\vartheta),
\ea
\ba
\label{e19}
\delta\tilde{v}(\vartheta,t)=\delta v_0\!-\!\omega_0\!\!\!\sum_{n=1}^{\infty}\!A_n\sin(\omega_n t+\varphi_n)\sin(n \vartheta),
\ea
where $\omega_n=n\omega_0$ denotes the quantized excitation energy. This equidistant excitation energy is also consistent with the $\mathrm{SU}(1, 1)$ dynamic symmetry.
The constant $A_0$ as a perturbation actually implies a small derivation from the chemical potential.
The constant $\delta v_0$ is set to be 0 with a Galilean transformation.

In the above calculation,  we assumed that the spatial derivative part of Gross-Pitaevskii (GP) equation can be neglected. The validity of this assumption has to be checked. By GP equation, we mean $\Phi(x,t)$ satisfy a non-linear Schr\"{o}dinger equation as follows,
\ba
i\partial_t\Phi=\!\left(\!-\frac{\partial_x^2}{2}\!-\!\mu\!+\!\frac{\omega_0^2x^2}{2}\!+\!\frac{g_0^2\pi^2}{2}|\Phi|^4\!\right)\Phi,
\ea
where $\mu$ denotes the chemical potential, $\Phi$ is short for $\Phi(x,t)$.
A direct numerical calculation for the excited states from the GP equation is difficult.
Here we carry out the check in a reversed way. We set the initial wave function as a superposition of the ground state and the $n$-th excited mode.
Then, we check the time evolution of the density wave with this initial condition,  where the information of the excited modes can be extracted.  Approximately, we  take the initial state as
\ba
\Phi_n(x, 0)=\sqrt{\rho_g(x)+0.1\cos[n\arccos(x/R)]}.
\ea
The dynamics of $\rho_n(x,t)\equiv |\Phi_n(x,t)|^2$ can be calculated by the GP equation. We find the dynamics of $\rho_n(x,t)$ can be factorized into several harmonic oscillations,
\ba
\delta\rho_n(x,t)&\equiv& \rho_n(x,t)-\overline{\rho_n(x,t)}\nonumber\\
&=&\sum_{m}A_m(x)\cos[\omega_m(x) t+\varphi_m(x)],
\ea
where $\overline{\rho_n(x,t)}=\int_0^{T_{\text{max}}}\mathrm dt\rho_n(x,t)/T_{\text{max}}$ is the average value of $\rho_n(x, t)$. $T_{\text{max}}$ is the time cutoff, which is taken as $50\pi$ here.
These data are shown in Fig.~\ref{Exs}. By a nonlinear fitting of the dynamics, we obtain the values of $A_m(x)$, $\omega_m(x)$ and $\varphi_m(x)$. These parameters are $x$ dependent.
If our result matches the theoretical predictions, $\omega_m(x)=\omega_m$ is $x$ independent.
Furthermore, we check that $\omega_m$ is interaction $g_0$ independent. This property is exact in $N\rightarrow\infty$ limit, but it is approximate for finite $N$.
Also, we check $\omega_m= m\omega_1$. If this equation fails to be true, there are soft modes in the system. The existence of these soft modes will lead to decoherence for the long-time evolution.
As we find $\tilde{\omega}_m(x)=\omega_m(x)/m$ are all very close to $1$, therefore we will give the mean value of $\omega(x)$ as $\overline{\omega}_m=\int_{-X_{\rm max}}^{X_{\rm max}}\mathrm dx \tilde{\omega}_m(x)/(2X_{\rm max})$
and its variance $\Delta \omega_n^2$ as a function of $g_0$ and $N$ in Tables~\ref{ta1}, \ref{ta2}. In Fig.~\ref{Exs}, one can observe that the dynamics in (b) is more stable than in (a), this is manifested in the frequency difference between $\bar{\omega}_2$ and $\bar{\omega}_4$. The difference between $\bar{\omega}_2$ and $\bar{\omega}_4$ shrinks from $0.003$ to $0.0002$ when the particle increase from $N=40$ to $N=125$.

\begin{table}
\centering

\caption{In this table, we show the $\bar{\omega}_{n=2,4,6}$ for different excited modes. The particle number $N$ is fixed at $40$. The average is over different positions. $\Delta \omega_n^2$ are the variance of $\bar{\omega}_n(x)$.  \label{ta1}}
\begin{tabular}{|c|c|c|c|c|c|c|}
\hline\hline
$N=40$&$\ \ \ \bar{\omega}_2$\ \ \ &$\ \ \Delta \omega_2^2$\ \ &\ \ \ $\bar{\omega}_4\ \ \ $&\ \ $\Delta \omega_4^2$\ \ &\ \ \ $\bar{\omega}_6$\ \ \ &\ \ $\Delta \omega_6^2$\ \ \\
\hline
$g_0=0.40$&$1.01003$&$0.0006$&$0.99752$&$0.0005$&$1.00114$&$0.0065$\\
\hline
$g_0=0.60$&$1.01958$&$0.0200$&$1.00379$&$0.0005$&$0.98812$&$0.0184$\\
\hline
$g_0=0.80$&$1.01109$&$0.0005$&$1.00558$&$0.0002$&$0.98496$&$0.0233$\\
\hline
$g_0=1.20$&$1.01307$&$0.0017$&$1.00675$&$0.0065$&$0.99835$&$0.0002$\\
\hline
$g_0=1.40$&$1.01468$&$0.0011$&$1.01045$&$0.0040$&$1.00021$&$0.0010$\\
\hline
\end{tabular}
\end{table}

\begin{table}
\centering

\caption{In this table, we show the $\bar{\omega}_{n=2,4,6}$ for different excited modes. The particle number $N$ is fixed at $125$. The average is over different positions. $\Delta \omega_n^2$ are the variance of $\bar{\omega}_n(x)$. \label{ta2}}
\begin{tabular}{|c|c|c|c|c|c|c|}
\hline\hline
$N=125$&$\ \ \ \bar{\omega}_2$\ \ \ &$\ \ \Delta \omega_2^2$\ \ &\ \ \ $\bar{\omega}_4\ \ \ $&\ \ $\Delta \omega_4^2$\ \ &\ \ \ $\bar{\omega}_6$\ \ \ &\ \ $\Delta \omega_6^2$\ \ \\
\hline
$g_0=0.40$&$1.00568$&$0.0046$&$1.00504$&$0.0001$&$0.99643$&$0.0025$\\
\hline
$g_0=0.60$&$1.00031$&$0.0228$&$1.01004$&$0.0190$&$1.00122$&$0.0015$\\
\hline
$g_0=0.80$&$1.00381$&$0.0094$&$1.00743$&$0.0010$&$0.99953$&$0.0066$\\
\hline
$g_0=1.20$&$1.00442$&$0.0032$&$1.00371$&$0.0010$&$1.00161$&$0.0010$\\
\hline
$g_0=1.40$&$1.00383$&$0.0025$&$1.00371$&$0.0012$&$0.99702$&$0.0020$\\
\hline
\end{tabular}
\end{table}

\begin{figure}[b]
\centering
\includegraphics[width=8cm]{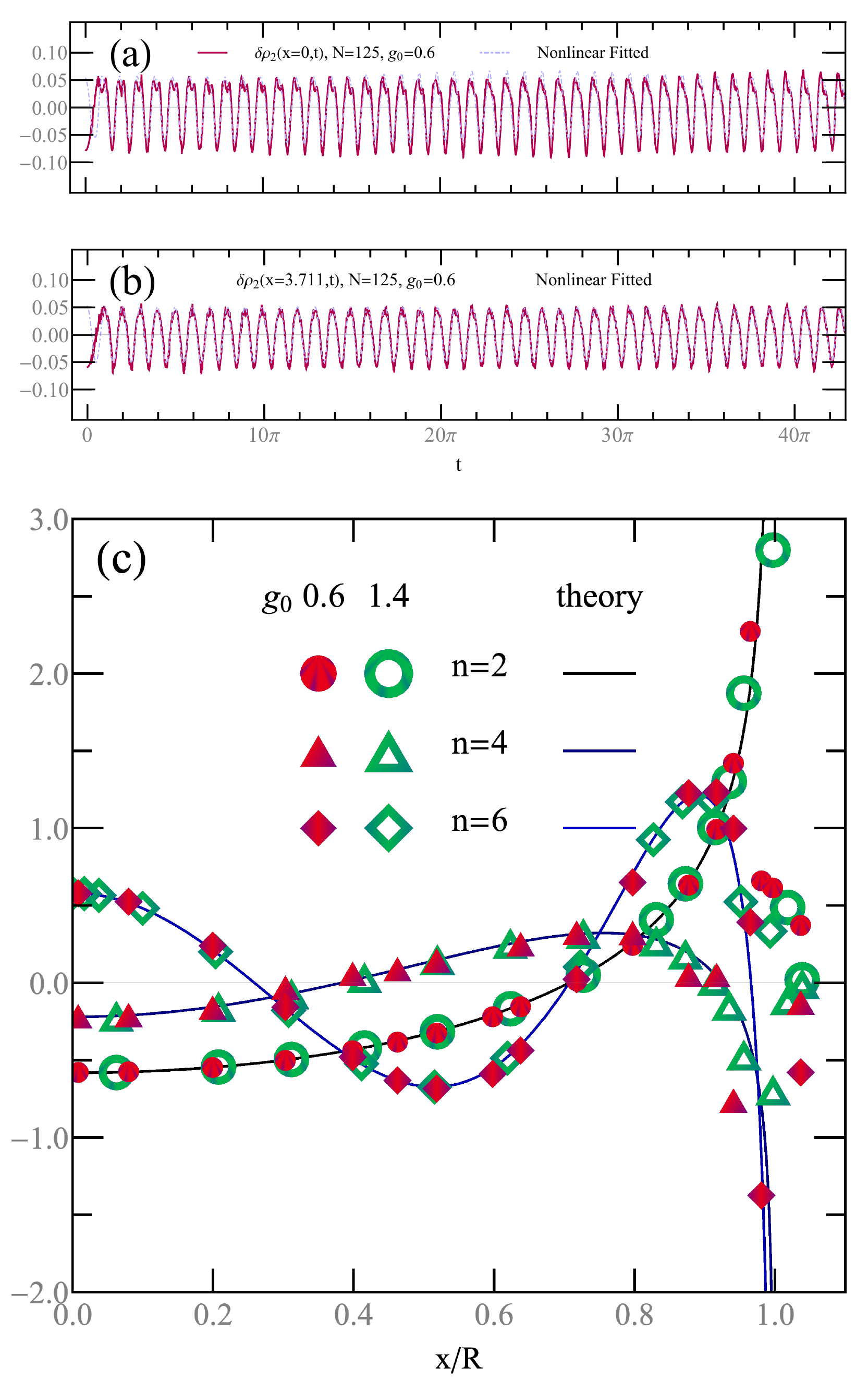}
\caption{In (a) and (b), we show two typical time evolution of the excited modes   for $\delta\rho_2(x,t)$ at different positions $x=0$ and $x=3.711$. $N=125$ is fixed. By fitting the curves, we  obtain $A_2(x)$, $A_4(x)$ and $A_6(x)$, which are even function of $x$.
By fixing the value of $A_{n=2,4,6}(x=0)$ to be the same, we give the $A_{n=2,4,6}(x/R)$ in (c). 
}
\label{ExPhi}
\end{figure}

We also study the $x$ dependence in $A_m(x)$, which gives us the information about the eigenfunctions of the excited modes,
\ba
A_m(x)\propto \frac{\delta\tilde{\rho}_m(x)}{\sqrt{1-(x/R)^2}}.
\ea
We find these relations are satisfied both for $N=40$ and $N=125$. In Fig.~\ref{ExPhi}(a) and (b), we show two typical $\delta\rho_2(x,t)$ at two different positions $x=0$ and $x=3.711$ ($x/R=0.3$).
In Fig.~\ref{ExPhi} (c), we show $A_2(x)$, $A_4(x)$ and $A_6(x)$ for $g_0=1.4$ and $g_0=0.6$ with the reference of theoretical prediction of $\delta\tilde{\rho}_2(x)/\sqrt{1-(x/R)^2}$ and $\delta\tilde{\rho}_4(x)/\sqrt{1-(x/R)^2}$ ($A_m(x)$ are even function of $x$).
One finds precise agreement  between the numerical results and the theoretical predictions when $|x|<R$. When $|x|\approx R$, the derivation indicates the finite-$N$ effects.
The eigenfunctions show $g_0$ independence up to a rescaling of $R$ ($R$ is in general $g_0$ dependent).
\begin{figure*}[t]
\centering
\includegraphics[width=15cm]{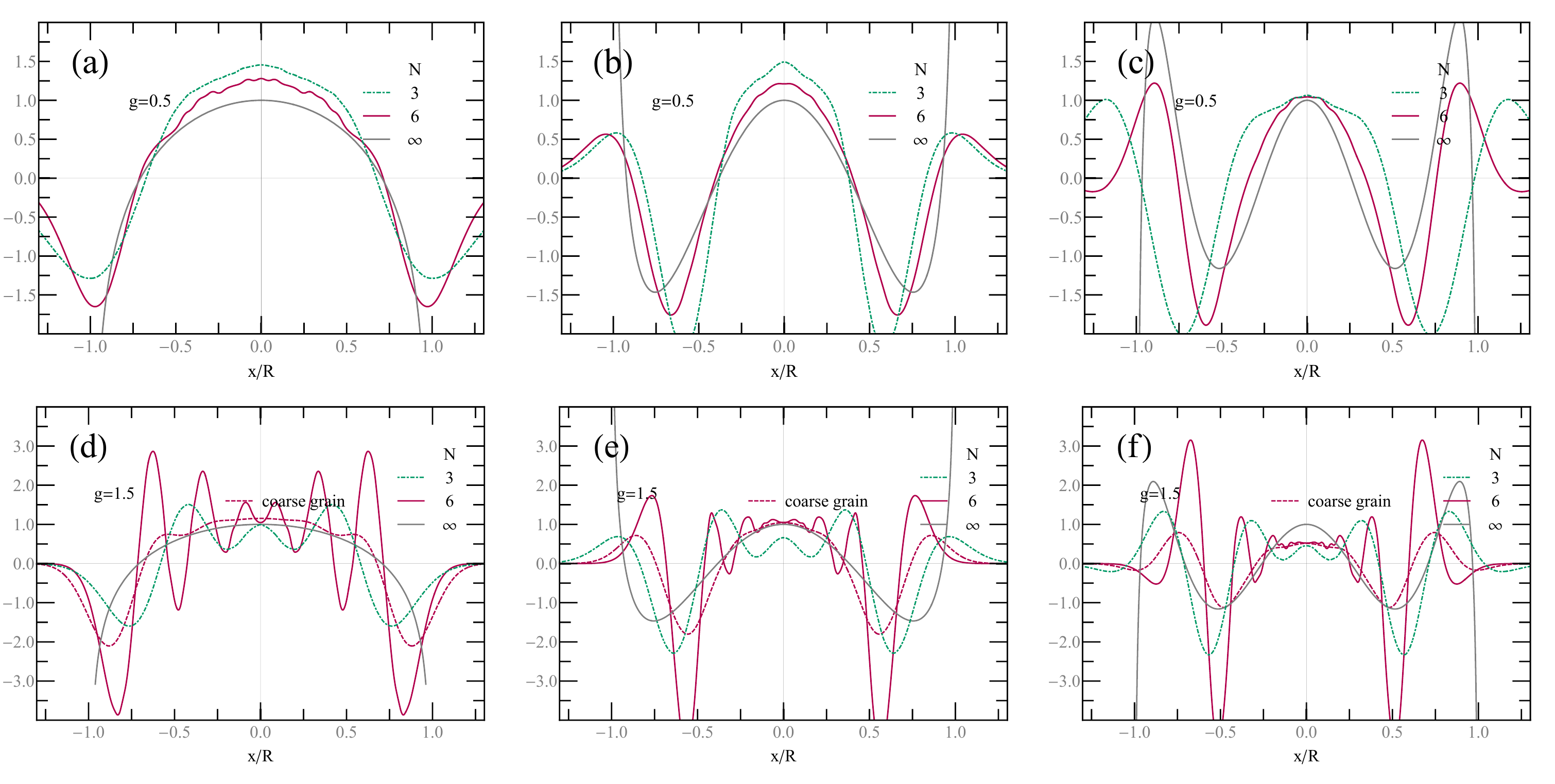}
\caption{In (a), (b), and (c), we show the density-wave excitation $\delta\rho_2^{\rm CS}(x)$, $\delta\rho_4^{\rm CS}(x)$ and $\delta\rho_6^{\rm CS}(x)$ for fixed $g_{\rm cs}=0.5$, respectively.
The system size is taken as $N=3$ ( the dot-dashed line ) and $N=6$ ( the purple solid line ). The reference $N=\infty$ line is given $\delta\rho_{n}^{\infty}(x)=\cos[n\arccos(x/R)]/\sqrt{1-(x/R)^2}$.
In (d), (e), and (f), the density wave excitation $\delta\rho_2^{\rm CS}(x)$, $\delta\rho_4^{\rm CS}(x)$ and $\delta\rho_6^{\rm CS}(x)$ are given for  $g_{\rm cs}=1.5$, respectively.
The purple dashed lines in (d), (e), and (f) are the coarse graining of $N=6$ curves.  }
\label{ExCSM}
\end{figure*}

Secondly, we study the density-wave excitation  of the CS model. Because  the CS model  can be mapped identically to a set of free harmonic oscillators with $E_g$ as the ground-state energy~\cite{gurappa1999}, the excited energy is also equidistant.
We consider the first excited state in the irreducible subspace involving the ground state, that is~\cite{csm1969, gambar1975}
\ba
\Psi_1(\{x_i\})=\left(\frac{E_g}{ \omega_{\rm cs}}\right)^{-1/2}\left(\frac{E_g}{\omega_{\rm cs}}-\omega_{\rm cs} \sum_{i=1}^{N}x_i^2\right)\Psi_{\rm g} .
\ea
More generally, we have
 \begin{gather}
  \begin{split}
\Psi_m(\{x_i\})=&\!\sqrt{mB\left(\frac{E_g}{\omega_{\rm cs}},m\right)}L_m^{\frac{E_g}{\omega_{\rm cs}}\!-\!1}\! \left(\omega_{\rm cs} \sum_{i=1}^{N}x_i^2\right)\\
&\times \Psi_{\rm g} (\{x_i\}),
    \end{split}
 \end{gather}
where 
the index $m$ in $\Psi_m$ refers to the $m $-th excited state in the subspace. 
Here $B(p,q)$ denotes the Beta-function, and $L_m^{\alpha}(x)$ denotes the generalized Laguerre polynomials.
Notice that the $j$-th excited state's energy is $E_g+2j\omega_{\rm cs}$.

Let $\Psi(  t=0)=c_0\Psi_{\rm g}  +c_1\Psi_1 $ with $|c_0|^2+|c_1|^2=1$. We  extract the information from the time evolution of the density
$\rho_1^{\rm CS}(x,t)=\int \prod_{j=2}^N \mathrm dx_j|\Psi(  t)|^2=\int \prod_{j=2}^N \mathrm dx_j[|c_0|^2\Psi_{\rm g}^{ 2}+|c_1|^2\Psi_1^2+(c_0c_1^*e^{2i\omega_{\rm cs} t}+c_0^*c_1e^{-2i\omega_{\rm cs} t})\Psi_{\rm g}\Psi_1]$. 
Since $\delta\rho$ is time dependent, we have $\delta\rho_1^{\rm CS}(x,t)=(c_0c_1^*e^{2i\omega_{\rm cs} t}+c_0^*c_1e^{-2i\omega_{\rm cs} t})\delta\rho_1^{\rm CS}(x)$, where
\begin{eqnarray}
&&\delta\rho_1^{\rm CS}(x)=\int_{-\infty}^{\infty}\prod_{j=2}^{N}\mathrm dx_j \Psi_{\rm g} (x, \{x_j\})\Psi_1(x, \{x_j\})\nonumber\\
          &=&\int_{-\infty}^{\infty}\prod_{j=2}^{N}\mathrm dx_j C\left(\frac{E_g}{\omega_{\rm cs}}-\omega_{\rm cs}\sum_{i=1}^{N}x_i^2\right)C^{-1}\Psi_{\rm g}(x, \{x_j\})^2\nonumber\\
          &=&\int_{-\infty}^{\infty}\prod_{j=2}^{N}\mathrm dx_j C\left(\frac{E_g}{\omega_{\rm cs}}+\omega_{\rm cs}\partial_{\omega_{\rm cs}}\right)\left[C^{-1}\Psi_{\rm g}(x, \{x_j\})^2\right]\nonumber\\
          &=&C\left(\frac{E_g}{\omega_{\rm cs}}-\omega_{\rm cs}\frac{\partial}{\partial\omega_{\rm cs}}\right)\left[C^{-1}\rho_{\rm g}^{\rm CS}( x)\right]\nonumber\\
          &\propto&\frac{\cos [2\arccos(x/R)]}{\sqrt{1-(x/R)^2}},
\end{eqnarray}
which is consistent with Eq.~(\ref{e18}) for $n=2$. In the derivation, we have assumed that the partial derivative $\partial_{\omega_{\rm cs}}$  and the thermodynamic limit are interchangeable.
However, for higher excited states which involves higher partial derivatives, this assumption is invalid due to the divergent derivative of $\rho_{\rm g}^{\rm CS}( x)$ at the boundary $x=\pm R$.
Although we can not prove the general form of $\delta\rho_m(x)$,  we can guess its form is consistent with Eq.~(\ref{e18}), i. e.,
\ba
\delta\rho_m^{\rm CS}(x)&=&\int_{-\infty}^{\infty}\prod_{j=2}^{N}\mathrm dx_j \Psi_{\rm g} (\{x_j\})\Psi_m(\{x_j\})\nonumber\\
&\propto& \frac{\cos[2m\arccos(x/R)]}{\sqrt{1-(x/R)^2}}\label{Guess}
\ea

For an arbitrary $g_{\rm cs}$, it is too difficult to solve the overlap integral for an any excited state, but Eq.~(\ref{Guess}) can be  proved when $g_{\rm cs}=1$.
Here, one of the  excited states   can be represented by the single-particle energy eigenstates of the harmonic oscillator $\{f_n(x)\}_{n=0,1,2,\cdots.}$.
Let $|0,1,\cdots, N-1\rangle$ denote the occupied single-particle energy eigenstates in the ground state of $N$ fermions.
Then, one of the $m $-th excited states $|\psi_m\rangle=\hat A\hat b^\dag_{N-l+m}\hat b_{N-l}|0, 1,\cdots, N-1\rangle$, for some $1\leq l\leq m$, where  $\hat A$ denotes the unit antisymmetric function~\cite{girardeau1960},
$\hat  b^\dag_l$ ($\hat b_l$) denotes the creation (annihilation) operator of $l$ single-particle energy eigenstates.  The corresponding perturbation of the density is given by (notice $\hat A^2=1$)
\ba
\delta\rho_m^{\rm CS}(x)&=&\int_{-\infty}^{\infty}\prod_{j=2}^{N}\mathrm dx_i\psi_0(x,\{x_j\})\psi_m(x,\{x_j\})\nonumber\\
&=&f_{N-l}(x)f_{N-l+m}(x).
\ea
Using the asymptotic expansion of the product in the thermodynamic limit (appendix A):
\begin{gather}
 \begin{split}
 \label{ea}
&f_{N-l}\left(\sqrt{\frac{2N}{\omega_{\rm cs}}}x\right)f_{N-l+m}\left(\sqrt{\frac{2N}{\omega_{\rm cs}}}x\right)\\
&\sim \sqrt{\frac{\omega_{\rm cs}}{2\pi^2N}}\frac{\cos(m\arccos x)}{\sqrt{1-x^2}}.
 \end{split}
\end{gather}
Comparing this result with Eq.~(\ref{e18}), we prove that the density-wave excitations in the trapped $|\Phi|^6$ theory and the CS model are the same when $g_{\rm cs}=1$.

Now, we are going to check  Eq.~(\ref{Guess}) for finite-$N$ systems by numerics. The overlap integral is numerically done by using Monte-Carlo method because  the integrant is semi-positive definite. Here we introduce a normalization condition $\int \mathrm dx \delta\rho_m(x)^2\rho_{\rm g}^{\rm CS}(x)=\pi/2$ for the comparison between different results, where $\delta\rho_m(x)$ and $\rho_{\rm g}^{\rm CS}(x)$ are numerically obtained with a finite $N$. 
In Fig.~\ref{ExCSM}, we present $\delta\rho_{m=1,2,3}(x)$ for $g_{\rm cs}=0.5$. There are datas for two sizes: $N=3$ and $N=6$. One can see the tendency for $\delta\rho_m(x)$ approaching the guessed function in the thermodynamic limit. The results for $g_{\rm cs}=1.5$ are given in (d), (e) and (f). There are clear deviations from the guessed results due to the high-frequency oscillation. The coarse-graining line is taken by replacing $\delta\rho_m(x)$ as $\int_{x-\Delta x}^{x+\Delta x}\mathrm dx'\delta\rho_m(x') /(2\Delta x)$.
Here $\Delta x$ is taken as $0.25$. The purple dashed lines are results after two times of coarse graining, which can be viewed as the average behavior of $\delta\rho_m(x)$. One can see their resemblance with the guessed result. From the numerical result we find when $g_{\rm cs}\gg 1$, then the deviation from the theoretical prediction is more serious. In the CS model, when $g_{\rm cs}$ is large, the particles are inclined to be aligned in equal space in
one-dimension due to the long-range   interaction. On the other hand, when $g_{\rm cs}< 1$, 
$\vartheta=\arccos(x/R)$ characterizes the oscillation of the density wave. 
But the oscillation in $\vartheta$ coordinate  is not consistent with the equal spacing in $x$ coordinate. Thus these two requirements are conflicting with each other.
For this reason, when $g_{\rm cs}/N\rightarrow 0$   is a suitable requirement for our guess. 

\subsection{Fractional exclusion statistics in $|\Phi|^6$ theory}

The wave function Eq.~(\ref{e16}) implies the canonical quantization of the density-wave excitation of the trapped $|\Phi|^6$ theory. Following the procedure of the phenomenological bosonization in Ref.~\cite{giamarchi2003},
we introduce a scalar field $\phi$ and its conjugate momentum $\Pi$ through $\delta \tilde{\rho}=-\partial \phi/\partial \vartheta,\delta \tilde{v}=\omega(g\pi)^{-1}\Pi$. Then, using the solution (Eqs.~\ref{e18}, \ref{e19})
and to the second order of $\delta\rho$, the energy functional (Eq.~\ref{energy}) is rewritten in terms of $\phi$ and $\Pi$ as
\ba
\label{e22}
H=\frac{1}{2}N^2g_0\omega_0+\frac{1}{2}\int_{0}^{\pi}\mathrm d\vartheta \left[\frac{\omega_0}{g_0\pi}\Pi^2+\pi g_0\omega_0\left(\frac{\partial \phi}{\partial \vartheta}\right)^2\right],
\ea
which describes a Luttinger-liquid system. In the derivation, we neglect the term $ \int \mathrm dx(\partial \rho_0/\partial x)^2$ within the TF approximation.

We quantize the scalar field with the commutation relation $[\hat \phi(t,y),\hat\Pi(t,y')]=i\delta(y-y')$. Thus, the density wave (Eq.~\ref{e18}, \ref{e19}) is of bosonic-excitation type. 
It follows from Eqs.~(\ref{e18}), (\ref{e19}) that
\ba
\label{e23}
\hat H=\frac{1}{2}g_0N^2\omega_0+\sum_{n=1}^{\infty}n\omega_0 \hat a_n^\dag\hat a_n,
\ea
where $\partial \hat\phi/\partial  \vartheta=\sqrt{2/\pi}\sum_{n}(2g_0\pi n)^{-1/2}(\hat a_n^\dag+\hat a_n)\cos(n\vartheta)$, $\hat\Pi=\sum_{n}i(g_0\pi n/2)^{1/2}(\hat a_n^\dag-\hat a_n)\sqrt{2/\pi}\sin(n\vartheta)$ with $\{\hat a_{n}^\dag,\hat a_n\}$ are the creation and annihilation operators of the density-wave modes.
Compared with the energy spectrum of the CS model~\cite{gurappa1999}, Eq.~(\ref{e23}) is also a set of free harmonic oscillators with $\frac{1}{2}g_0N^2 \omega_0$ as the ground-state energy.
Because, the only difference between the two models is just a negligible chemical potential deviation $(1-g_0)\omega_0/2$ for $\omega_0\to 0$ in the thermodynamic limit,
we conclude that the trapped $|\Phi|^6$ theory has the CS model-like energy spectra and thus obeys the fractional exclusion statistics with  $g_0$ as the statistical parameter.
About the connection between the Luttinger-liquid and the fractional exclusion statistics, we introduce Refs.~\cite{Pham1998, Wu2001} for further reading .


\begin{figure}[t]
\centering
\includegraphics[width=8.cm]{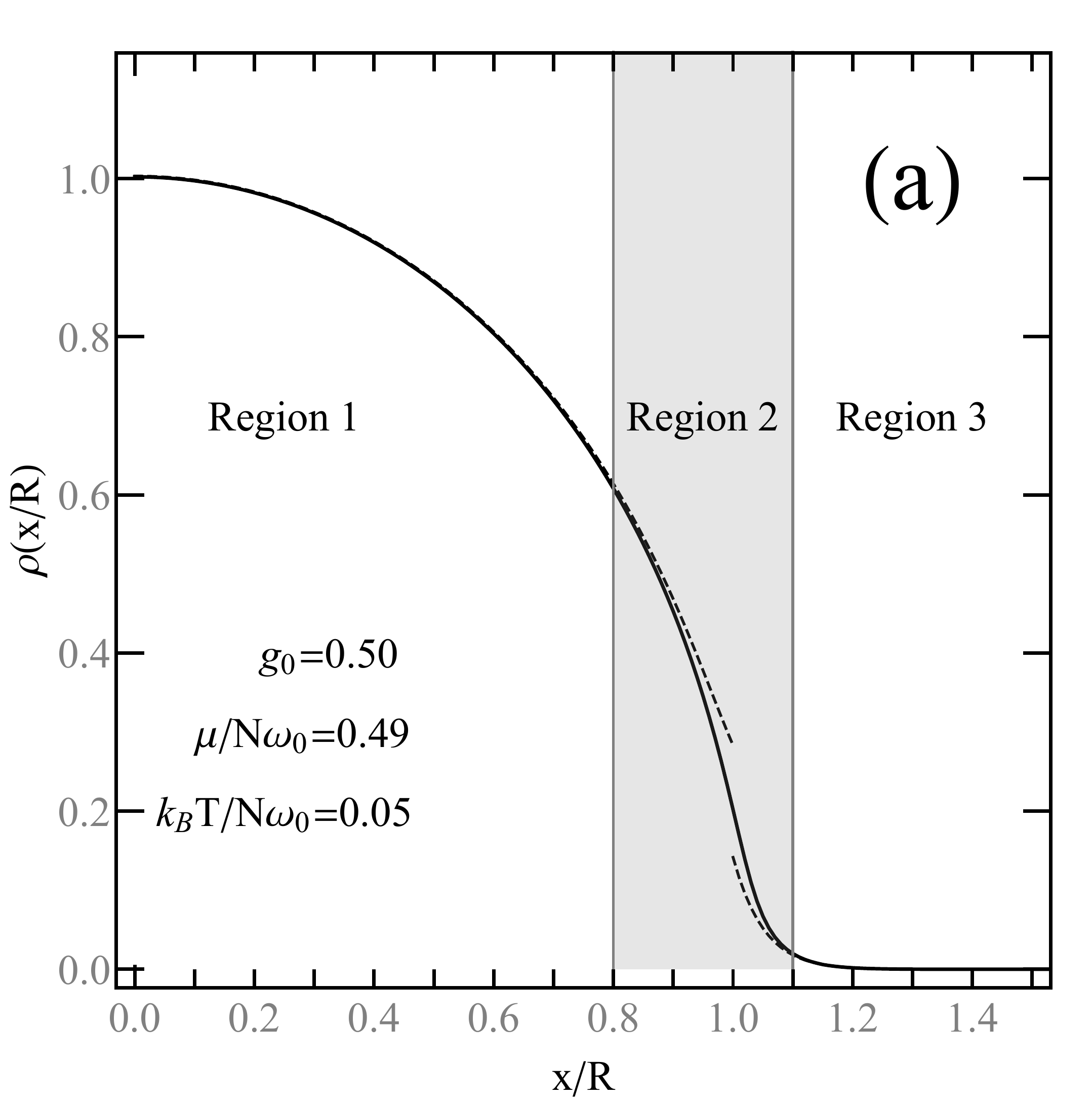}
\includegraphics[width=8.cm]{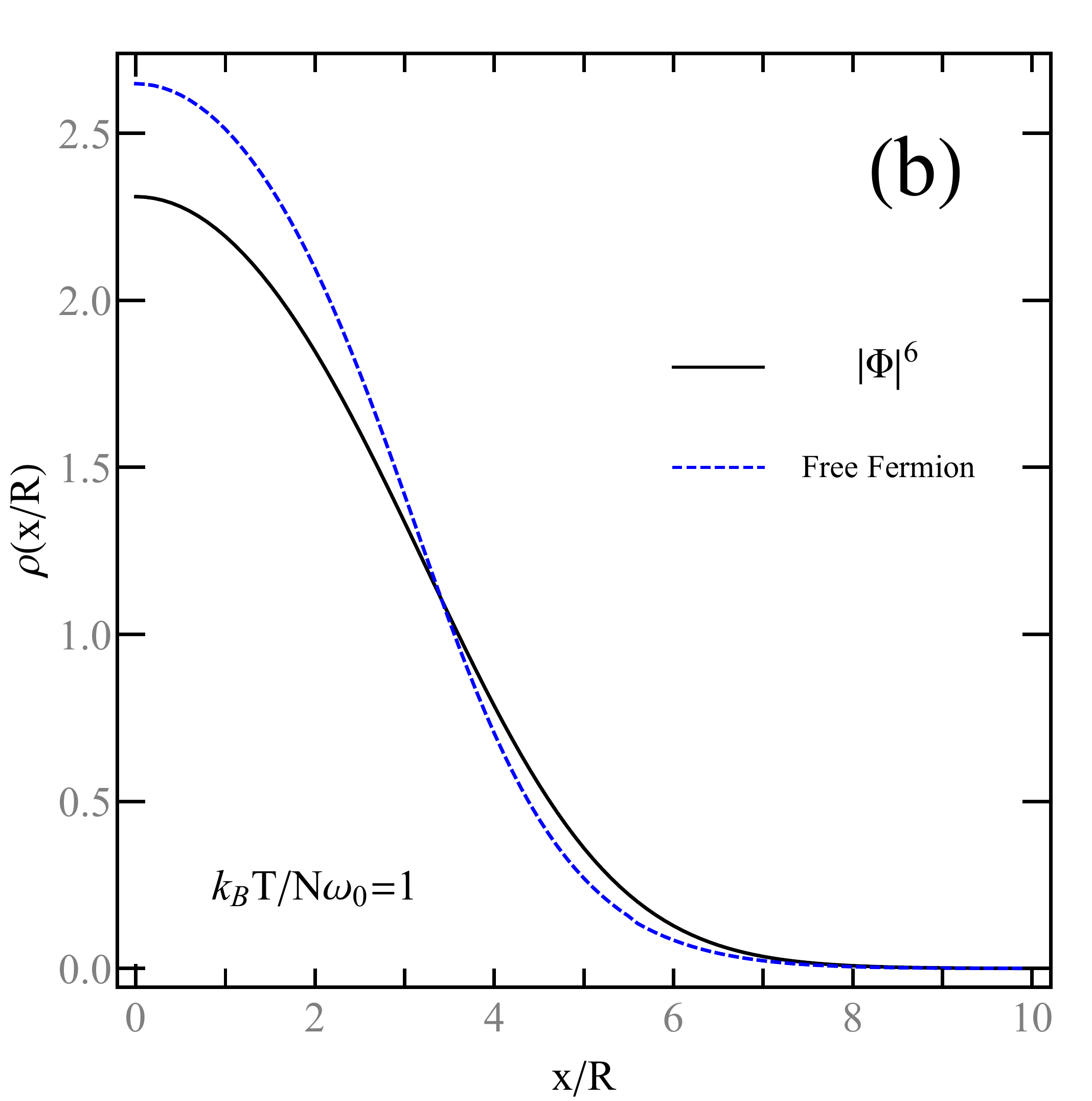}
\caption{The  density of the $|\Phi|^6$ theory at finite temperature. For $x<0$, the density is an even function. (a) The black solid line denotes the numerical result from Eq.~(\ref{e124}). The red and blue solid lines denote the approximate results from Eq.~(\ref{e27}) and Eq.~(\ref{e28}) respectively. (b) The black solid line denotes the numerical result from Eq.~(\ref{e124}). The blue dashed line denotes the  density of a ideal fermionic system with the same temperature.}\label{fig1}
\end{figure}

\subsection{Finite Temperature Results}

The linearized density-wave excitations capture the low-energy excitations of the $|\Phi|^6$ theory. When the temperature $k_{\text B} T$ ($k_{\text B}$ denotes the Boltzmann constant) of the field grows up to $N\hbar\omega$, the thermal-excitation energy in Eq.~(\ref{e23}) is comparable to the ground-state energy.
At this moment, the linearized theory fails and we use the semiclassical approximation to study the finite-temperature property of the $|\Phi|^6$ theory in the following.

Using the Hatree-Fock  approximation and the WKB approximation~\cite{huse1982,giorgini1997,bhaduri2000}, the density $\rho(x)$ is determined self-consistently. First there should no condensate part in $\rho(x)$ at a high temperature.
The system consists of thermal particles and the  density  $\rho(x)$ at finite temperature $\beta=1/k_{\text B} T$ reads
\ba
\label{e24}
\rho(x)=\int_{-\infty}^{\infty}\frac{\mathrm dp/(2\pi\hbar)}{e^{\beta\varepsilon(p,x)}-1},
\ea
where
\ba
\label{e25}
\varepsilon(p,x)=\frac{p^2}{2}+\frac{1}{2}\omega_0^2x^2+\frac{g_0^2\pi^2\hbar^2}{2}\rho(x)^2-\mu
\ea
denotes the effective single-particle spectrum, and   $\mu$ denotes the chemical potential. After performing the $p$-integration of Eq.~(\ref{e24}), we find the self-consistent equation of $\rho(x)$:
\ba
\label{e124}
u(x)=\mathrm{Li}_{1/2}[e^{\beta\mu-\beta \omega_0^2x^2/2-\pi g_0^2 u(x)^2/4}],
\ea
where $u(x)= \sqrt{2\pi\beta}\rho(x)$ denotes the dimensionless density, and $\mathrm{Li}_{s}(z)=\sum_{n=1}^{\infty}z^n/n^s$ denotes the polylogarithm function. Also, the constraint is given by
\ba
\label{e26}
N \sqrt{2\pi\beta}=\int_{-\infty}^{\infty}\mathrm dx u(x).
\ea
Specially, we find the solution of $\rho(x)$ can always be obtained self-consistently for any temperature, which indicates that there are no condensate in $\rho(x)$ at any  temperature.

\begin{figure}[t]
\centering
\includegraphics[width=8cm]{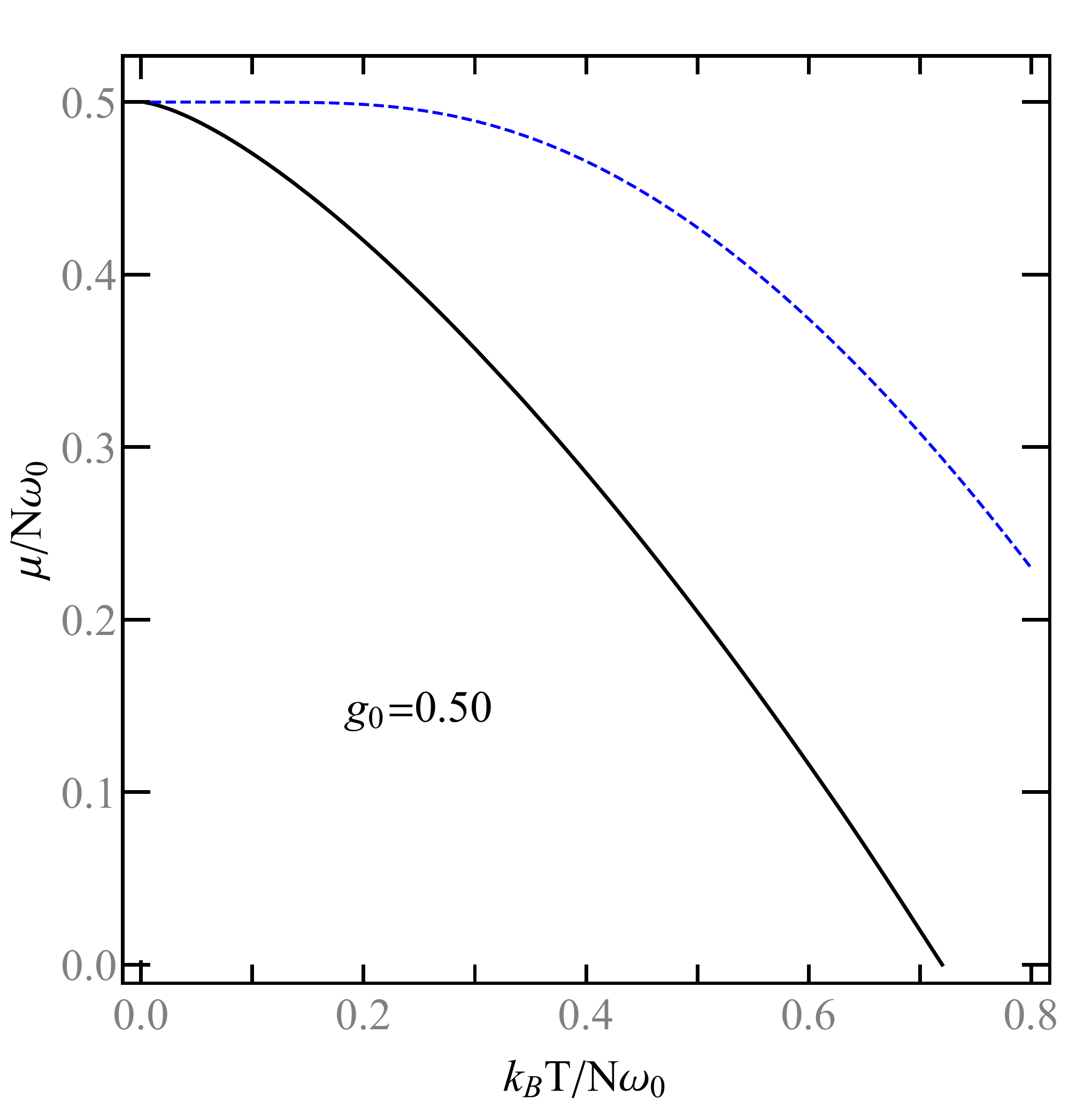}
\caption{The $\mu-T$ curve of the $|\Phi|^6$ theory at finite temperature. The black solid line denotes the numerical result from Eq.~(\ref{e124}).  The blue dashed denotes the $\mu-T$ curve of the CS model. }\label{fig2}
\end{figure}

In Fig.~\ref{fig1}, we show the density $\rho (x)$ numerically solved from Eq.~(\ref{e124}). Fig.~\ref{fig1}a  indicates three typical regions: (1) semicircle distribution: $|x|<R$; (2) Gaussian tail: $|x|>R$;
(3) near the boundary $|x|\approx R$. Actually, the region 1 and 3 respectively corresponds to the two limits of the polylogarithm function: $\mathrm{Li}_{1/2}(z)\to \sqrt{\pi/(1-z)}$ for $z\to 1^-$, 
and $\mathrm{Li}_{1/2}(z)\to z$ for $z\to 0$. Accordingly, the density  in the two regions are
\ba
\label{e27}
u^{1}(x)=\sqrt{\frac{(2\mu-\omega_0^2x^2)}{\pi g_0^2 k_{\text B} T}}\sqrt{1+\sqrt{1+\left(\frac{2\pi g_0 k_{\text B} T }{2\mu-\omega_0^2x^2}\right)^2}},
\ea
\ba
\label{e28}
u^{3}(x)=e^{\beta(\mu-\omega_0^2x^2/2)}.
\ea
Substituting Eq.~(\ref{e27}), or Eq.~(\ref{e28}) into Eq.~(\ref{e25}), we have
\ba
\varepsilon^{1}(p,x)=\frac{p^2}{2}+\frac{(\pi g_0 k_{\text B} T)^2}{4\mu-2\omega_0^2x^2}
\ea
which results in the semicircle distribution of the density and implies the existence of the Fermi surface, or
\ba
\varepsilon^{3}(p,x)=\frac{p^2}{2}+\frac{1}{2}\omega_0^2x^2-\mu
\ea
which describes the ideal Bosons, but now the chemical potential contains the effects of interaction. In addition, near the boundary $|x|\approx R$ with a length scale $R k_{\text B} T/\mu$, the analytical solutions (Eqs.~\ref{e27}, \ref{e28}) are invalid. 
Fig.~\ref{fig1}b compares the density of the $|\Phi|^6$ theory ($g_0=1$) with it of a free-fermion system for fixed $N$ and $T$. The narrower distribution of the density from the $|\Phi|^6$ theory indicates that more particles occupy the low-energy states, which lead to a lower chemical potential.

In Fig.~\ref{fig2}, we show the $\mu-T$ figure of the $|\Phi|^6$ theory (Eqs.~\ref{e124}, \ref{e26}) and the CS model~\cite{wu1994}
\ba
\label{e30}
\mu_{\text{CS}}=g_0N\omega_0+k_BT\ln[1-e^{-N\omega_0/k_BT}].
\ea
Compared with the CS model's result, the chemical potential of  the $|\Phi|^6$ theory drops rapidly as the temperature grows, which is consistent with Fig.~\ref{fig1}b.
The deviation of the two curves indicates that the $|\Phi|^6$ theory does not satisfy the fractional exclusion statistics at the temperature scale $k_{\text B} T\sim N\omega_0$,
in contrast to the result that the $|\Phi|^6$ theory  satisfies the fractional exclusion statistics within the linearized theory at the temperature scale  $k_{\text B} T\ll N\omega_0$.
This deviation is a result of the expansion of the semicircle distribution at finite temperature, which decreases the density and weakens the three-body interaction of the field (see also Fig.~\ref{fig1}b).

\section{Conclusion}

In this paper, we find a hydrodynamical description of ideal anyons satisfying fractional exclusion statistics in the one-dimensional $|\Phi|^6$ theory.
The $|\Phi|^6$ theory  shares many similar features as the Calogero-Sutherland model. Specifically, when the trap frequency and the interaction strength in $|\Phi|^6$ theory and them in the CS model are consistent
the two models share the same ground state density profile, excitation spectrum, density-wave excitation , as well as dynamical and statistical properties in the thermodynamic limit and at low temperature.
The hydrodynamical version of ideal  anyons satisfying fractional exclusion statistics is more accessible in experiments because it only relates local contact interactions, while $r^{-2}$ interaction in the CS model is, on the other hand, very difficult to realize experimentally.

\section{Acknowledgement}

Z. Y. Fei is supported by the NSFC (Grants No. 12088101), NSAF (Grants No. U1930403, No. U1930402), and  the China Postdoctoral Science
Foundation (No. 2021M700359). Y. Chen is supported by the National Key Research and Development Program of China (Grant No. 2022YFA1405300), and NSFC under Grant No. 12174358 and No. 11734010.

 \renewcommand{\theequation}{A.\arabic{equation}}

 \setcounter{equation}{0}

\section*{Appendix A: asymptotic expansion of $f_n$}

The eigenstates of the harmonic oscillator is given by
\ba
f_n(x)=\frac{1}{\sqrt{n!2^n}}\left(\frac{\omega_{\rm cs}}{\pi}\right)^{1/4}H_n(\sqrt{\omega_{\rm cs}}x)e^{-\frac{\omega_{\rm cs}}{2}x^2}.
\ea
In the thermodynamic limit, according to Ref.~\cite{Dominici2005}, we have ($|x|<1$)
\begin{gather}
 \begin{split}
f_N\left(\sqrt{\frac{2N}{\omega_{\rm cs}}}x\right)\sim  \left[\frac{2\omega_{\rm cs}}{\pi^2N(1-x^2)}\right]^{1/4}\cos\Theta_N(x),
 \end{split}
\end{gather}
where $\Theta_N(x)=(N+1/2)\arccos x-Nx\sqrt{1-x^2}-\pi/4$.
Then, for the product $f_{N-l}f_{N-l+m}$, in the limit $m/N\to 0$, we have
\begin{widetext}
\begin{gather}
 \begin{split}
&\sqrt{\frac{2\pi^2N(1-x^2)}{\omega_{\rm cs}}}f_{N-l}\left(\sqrt{\frac{2N}{\omega_{\rm cs}}}x\right)f_{N-l+m}\left(\sqrt{\frac{2N}{\omega_{\rm cs}}}x\right)\\
&\sim\frac{1}{2}\cos\Theta_{N-l}\left(\sqrt{\frac{N}{N-l}}x\right)\cos\Theta_{N-l+m}\left(\sqrt{\frac{N}{N-l+m}}x\right)\\
&=\cos\left[\Theta_{N-l}\left(\sqrt{\frac{N}{N-l}}x\right)-\Theta_{N-l+m}\left(\sqrt{\frac{N}{N-l+m}}x\right)\right]+\cos\left[\Theta_{N-l}\left(\sqrt{\frac{N}{N-l}}x\right)+\Theta_{N-l+m}\left(\sqrt{\frac{N}{N-l+m}}x\right)\right]\\
&\sim \cos( m\arccos x)+ \cos\left[\Theta_{N-l}\left(\sqrt{\frac{N}{N-l}}x\right)+\Theta_{N-l+m}\left(\sqrt{\frac{N}{N-l+m}}x\right)\right].
 \end{split}
\end{gather}
\end{widetext}
Ignoring the high frequency oscillating term (the second term), we obtain Eq.~(\ref{ea}).

\end{document}